\documentclass[letterpaper]{article}
\usepackage[preprint]{aaai2027}
\usepackage[hyphens]{url}
\usepackage{graphicx}
\usepackage{natbib}
\usepackage{caption}
\usepackage{booktabs}
\usepackage{amsmath}
\usepackage{amssymb}
\usepackage{multirow}
\usepackage{float}
\usepackage{xcolor}
\title{\begin{tabular}{@{}c@{}}
FRAUDSkill: Structured Frozen-Weight Skill Optimization\\
for Audio Anti-Fraud Detection
\end{tabular}}
\author{
    Chengxian Hu\textsuperscript{1}\equalcontrib, Zhiming Ma\textsuperscript{2}\equalcontrib, Mingjun Pan\textsuperscript{6}, Yifan Wang\textsuperscript{1} \\
    Shun Zhang\textsuperscript{1}\corresponding, Qifan Wang\textsuperscript{3}, Zhilei Zhao\textsuperscript{1}, Yijin Zhou\textsuperscript{4} \\
    Yuxi Zhao\textsuperscript{1}, Huiyuan Liu\textsuperscript{1}, Peidong Wang\textsuperscript{5}, Peng Chen\textsuperscript{1}
}
\affiliations{
    \textsuperscript{1}People's Public Security University of China \quad
    \textsuperscript{2}JD Technology \quad \textsuperscript{3}MetaAI \\
    \textsuperscript{4}University of Science and Technology of China \quad
    \textsuperscript{5}Northeastern University \\
    \textsuperscript{6}Peking University, Shenzhen \\[0.5ex]
    {\small
    \{202421430031, 2024111026, 202421450003, 202421250002\}@stu.ppsuc.edu.cn \\
    \{shunzhang, zhaozl, chenpeng\}@ppsuc.edu.cn \quad mazhiming312@outlook.com \\
    wqfcr@meta.com \quad zyjm@mail.ustc.edu.cn \quad pdongwang@163.com \\
    mingjunpan96@gmail.com}
}

\begin{document}
\maketitle

\begin{abstract}
Large audio-language models have shown promise for anti-fraud detection by directly processing speech and reasoning over fraud-related evidence. Their deployment, however, requires predictions to follow a predefined label space and a structured decision protocol consisting of service-scenario identification, fraud detection, and conditional fraud-type classification. Existing fine-tuning and prompt-based approaches typically encode task knowledge, constraints, and decision rules into model parameters or manually maintained prompts, making them difficult to adapt as fraud patterns and labeling policies evolve. To this end, we propose \textit{FRAUDSkill}, a structured frozen-weight adaptation framework that leaves the underlying audio-language model unchanged while optimizing an external layer of skill programs, route-specific policies, and decision rules. We further combine structured output control with validation-guided multi-path inference to ensure protocol-compliant predictions. On the TeleAntiFraud benchmark, FRAUDSkill achieves $73.50\%$ Macro-F1, outperforming the shared frozen-model baseline by $31.96\%$ while reducing invalid outputs to $1.94\%$. Extensive experiments demonstrate that external skill optimization provides an effective and adaptable solution for structured audio anti-fraud detection without modifying the underlying model. The source code is available at \url{https://anonymous.4open.science/r/FRAUDSKILL-114514}.
\end{abstract}

\section{Introduction}
Telecom fraud poses persistent threats to public safety, financial security, and social trust, creating an urgent need for automated detection from spoken interactions. Recent audio-language models  \cite{radford2023robust,chu2024qwen2audio,tang2024salmonn} provide a promising foundation for this task by directly processing speech and reasoning over fraud-related evidence. However, practical anti-fraud systems operate under predefined label ontologies, requiring predictions to follow a structured decision protocol consisting of service-scenario identification, fraud detection, and conditional fraud-type classification. Consequently, audio anti-fraud detection is not an open-ended understanding task, but a structured closed-set decision problem in which every prediction must satisfy application-specific constraints.

Recent advances in large language models (LLMs)  \cite{brown2020language} and multimodal large language models (MLLMs)  \cite{yin2024survey} have substantially expanded the use of foundation models for fraud detection across diverse domains, including scam-message detection  \cite{jiang2024detectingscams}, phone-scam analysis and real-time warning  \cite{shen2025combating,shen2025warned}, phishing and spam detection ~ \cite{jamal2024improved,blake2025phishsense}, payment-risk assessment  \cite{dahiphale2024payments}, and transaction-network fraud modeling  \cite{luo2025llmgcnfraud}. In the audio domain, recent benchmarks formulate telecom anti-fraud detection as structured prediction over service scenarios, fraud status, and fraud types  \cite{wang2026safeqaq}. These studies demonstrate the strong reasoning capabilities of large models, but practical deployment further requires predictions to conform to predefined label spaces and decision protocols enforced by real-world anti-fraud systems  \cite{ma2025teleantifraud}.

\begin{figure}[t]
  \centering
  \IfFileExists{figures/figure_teaser.png}{%
    \includegraphics[width=\columnwidth]{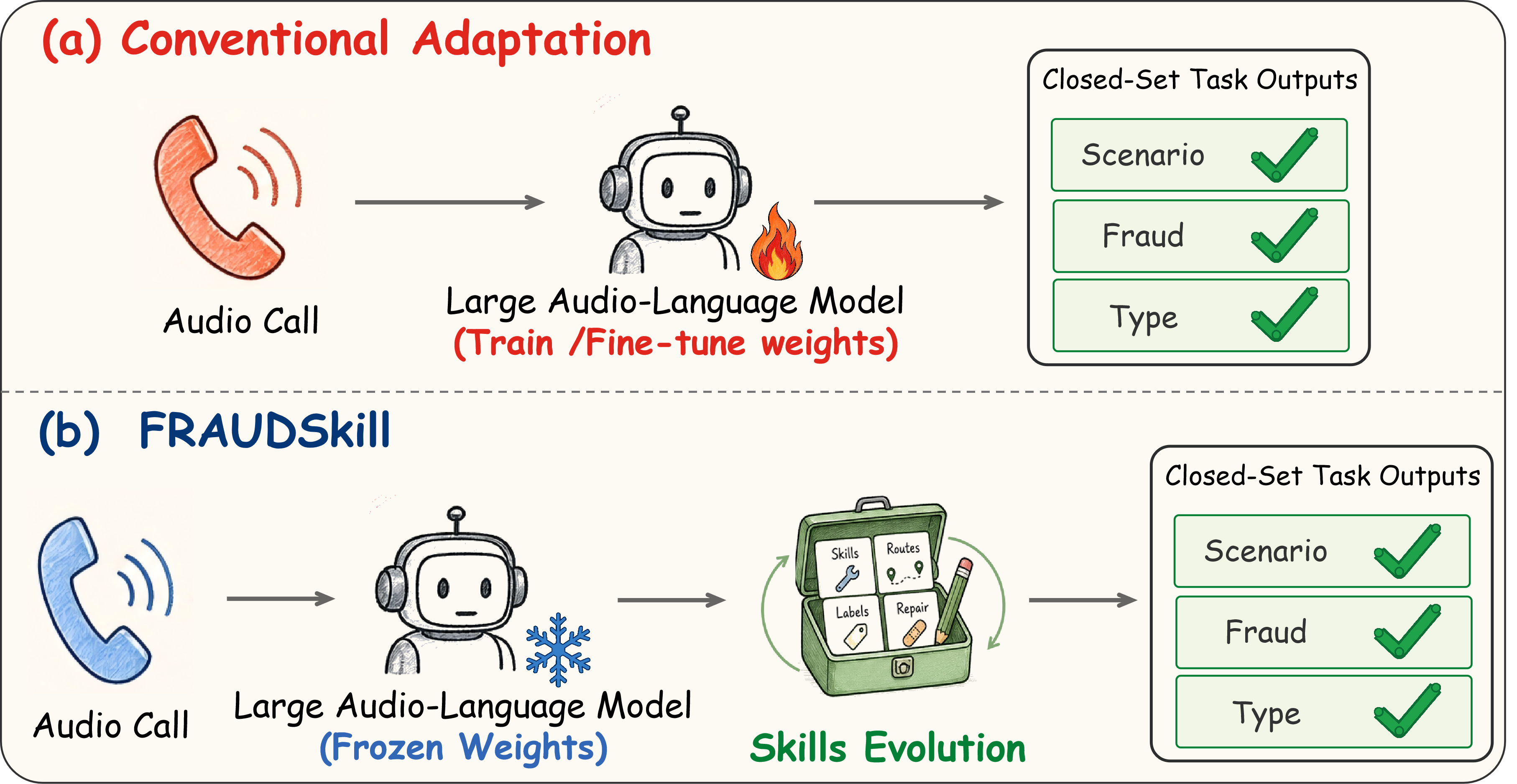}%
  }{%
    \fbox{\parbox[c][3.0cm][c]{0.94\columnwidth}{\centering Placeholder for the FRAUDSkill teaser figure.}}%
  }
  \caption{Comparison between conventional adaptation and FRAUDSkill. Conventional adaptation updates model weights or relies on manually revised instructions, whereas FRAUDSkill keeps the large audio-language model frozen and evolves an external skill layer for closed-set audio anti-fraud decisions.}
  \label{fig:teaser}
  \vspace{-3mm}
\end{figure}

In audio anti-fraud detection, closed-set predictions are inherently coupled through a hierarchical decision process. Given an audio input, the system first identifies the service scenario, then determines whether the interaction is fraudulent, and predicts a fraud type only when fraud is detected. This sequential dependency introduces challenges beyond semantic understanding: a prediction may be linguistically plausible yet violate the predefined label ontology, omit a required decision, or become inconsistent with earlier outputs  \cite{khattab2023dspy}. Since each stage conditions the next, an invalid early prediction can propagate through the pipeline and ultimately invalidate the entire decision process.

Existing adaptation approaches generally specialize foundation models by either updating model parameters or manually designing prompts  \cite{shin2020autoprompt,zhou2022large}. Parameter-efficient or full fine-tuning methods can achieve strong performance but implicitly encode task knowledge, label constraints, and decision policies into model weights, requiring additional optimization whenever fraud patterns, label definitions, or deployment policies evolve. Prompt-based methods are more flexible but often rely on manually engineered instructions that are difficult to maintain across models and provide limited guarantees for structured outputs and sequential decision consistency  \cite{pryzant2023automatic,fernando2024promptbreeder}. As a result, neither paradigm simultaneously offers adaptability, maintainability, and reliable enforcement of structured decision protocols while keeping the underlying model unchanged.

To address these challenges, we propose \textit{FRAUDSkill}, a structured frozen-weight adaptation framework for audio anti-fraud detection. Rather than modifying the audio-language model itself, FRAUDSkill learns an external skill layer from task examples. The skill layer explicitly represents fraud knowledge, label constraints, routing policies, and validation-derived decision preferences as modular, inspectable, and replaceable artifacts, enabling detection policies to evolve without retraining the underlying model.

FRAUDSkill is designed to align external skill optimization with the hierarchical and closed-set nature of audio anti-fraud detection. During optimization, it searches for route-aware skill programs that capture fraud evidence, task policies, and label constraints through program-based optimization  \cite{khattab2023dspy}. During inference, it combines structured output control with validation-guided multi-path inference to enforce valid labels, consistent routing, and protocol-compliant decision sequences. Together, these components enable reliable structured prediction while keeping the underlying audio-language model completely frozen.

Our main contributions are summarized as follows:
\begin{itemize}
    \item We formulate audio anti-fraud detection as a structured frozen-weight adaptation problem, where a frozen audio-language model must support controllable closed-set and chained detection decisions.
    \item We propose \textit{ FRAUDSkill}, which combines route-aware external skill optimization with structured multi-path inference, enabling task knowledge and decision policies to evolve without modifying the audio-language model.
    \item Experiments on the TeleAntiFraud benchmark show that \textit{FRAUDSkill} achieves $73.50\%$ Macro-F1, outperforming the shared frozen-model baseline by $31.96\%$ while reducing the invalid-output rate to $1.94\%$.

\end{itemize}

\section{Related Work}
\paragraph{LLMs and MLLMs for Anti-Fraud Detection.}

Anti-fraud detection has evolved from rule-based and task-specific systems to LLM- and MLLM-based approaches across text, speech, payment, and transaction-network settings \cite{terzi2021telecom}. More closely related to our setting, TeleAntiFraud-28k formulates spoken telecom fraud detection as structured prediction over service scenarios, fraud labels, and fraud types \cite{ma2025teleantifraud}, while SAFE-QAQ improves audio-text fraud reasoning through supervised fine-tuning and reinforcement learning \cite{wang2026safeqaq}. Although these studies demonstrate the effectiveness of large models for audio anti-fraud detection, their task adaptation primarily relies on updating model parameters. In contrast, our work investigates structured adaptation under a fully frozen audio-language model, externalizing task knowledge, label constraints, and decision policies into an optimizable skill layer.

\paragraph{Frozen-Model Prompt and Skill Adaptation.}

Frozen models can be adapted through prompt optimization, program or pipeline optimization, and external skill optimization. Prompt-level methods automatically search for or revise task instructions using discrete optimization, model feedback, or textual gradients \cite{yang2024large,fernando2024promptbreeder}. Program- and pipeline-level methods further optimize multiple model calls and intermediate components under task-level objectives \cite{yuksekgonul2024textgrad,agrawal2025gepa}. More closely related to our work, SkillOpt  \cite{skillopt2026} optimizes a unified external skill artifact, while EvoSkill  \cite{evoskill2026} discovers and revises modular skills according to execution feedback. Although SkillOpt and EvoSkill demonstrate that external skills can be optimized for frozen models, they remain task-general and do not explicitly support the conditional decision chains and closed-set deployment required by audio anti-fraud detection. FRAUDSkill is designed for this setting through a route-aware skill program that aligns external adaptation with the three-stage decision chain, official label constraints, and cross-route consistency, while keeping the underlying audio-language model frozen.

\section{Method}
\label{sec:method}
\begin{figure*}[t]
  \centering
  \IfFileExists{figures/figure_overview.png}{%
    \includegraphics[width=\textwidth]{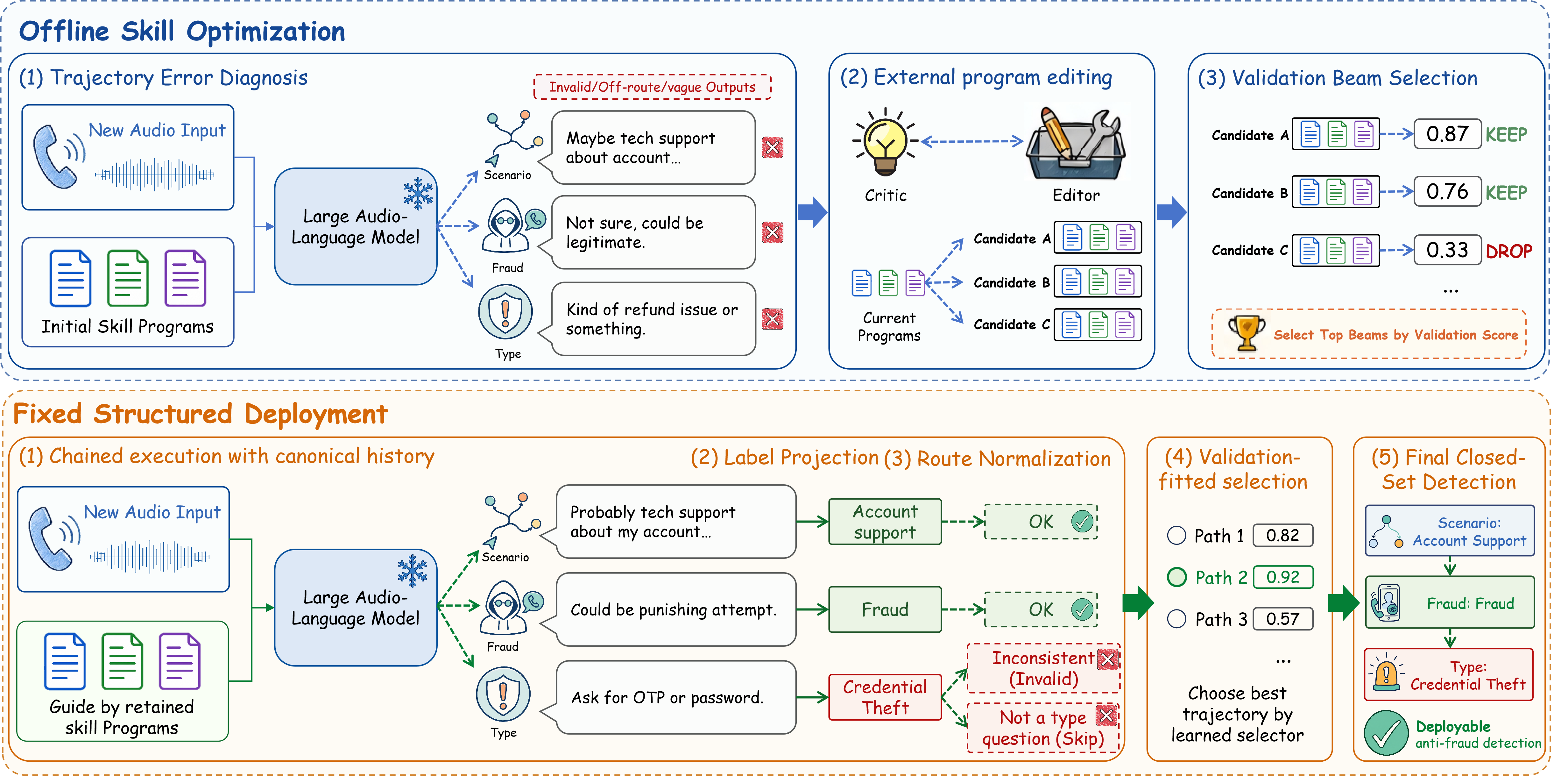}%
  }{%
    \fbox{\parbox[c][6.0cm][c]{0.96\textwidth}{\centering Placeholder for the FRAUDSkill framework figure.}}%
  }
  \caption{Overview of FRAUDSkill. Offline skill optimization edits only external skill programs and route policies using rollout errors, textual critique, and validation selection. The retained programs, label maps, route rules, and selector are then frozen as deployment artifacts. At test time, the frozen audio actor follows retained programs, produces multiple route trajectories, projects free-form outputs to official labels, repairs invalid chains, and selects a final closed-set anti-fraud decision.}
  \label{fig:overview}
\end{figure*}
\subsection{Problem Setting and Framework Overview}

FRAUDSkill adapts an audio anti-fraud system without updating the parameters of its audio-language model. The frozen actor still processes the audio and produces natural-language judgments. FRAUDSkill instead optimizes and deploys an external layer of task instructions, anti-fraud skills, routing policies, and structured decision rules. Its purpose is to convert the actor's open-ended audio understanding into closed-set predictions that satisfy the label ontology and decision protocol required by the deployed system.

Given an audio input $x$, the system makes three ordered decisions
$\mathcal{Q}=(\mathrm{scene},\mathrm{fraud},\mathrm{type})$.
The scene route predicts a service scenario, the fraud route predicts either \texttt{normal} or \texttt{fraud}, and the type route predicts a fraud category only when the fraud decision is \texttt{fraud}. Each route $q\in\mathcal{Q}$ has an official label set $\mathcal{Y}_q$ and a fixed route query $u_q$.

We distinguish an invalid output from a route that is correctly skipped. The symbol $\mathrm{INV}$ denotes a required output that is missing or cannot be mapped to an official label, whereas $\mathrm{NA}$ denotes a route that is not applicable under the protocol. The feasible output space is therefore
\begin{equation}
\label{eq:feasible_space}
\begin{aligned}
\mathcal{C}
&=
\mathcal{Y}_{\mathrm{scene}}\!\times\!\{\texttt{normal}\}\!\times\!\{\mathrm{NA}\} \\
&\quad \cup\,
\mathcal{Y}_{\mathrm{scene}}\!\times\!\{\texttt{fraud}\}\!\times\!\mathcal{Y}_{\mathrm{type}} .
\end{aligned}
\end{equation}
Required outputs mapped to $\mathrm{INV}$ are outside $\mathcal{C}$ and are counted as errors.

FRAUDSkill has two strictly separated stages. During \emph{offline skill optimization}, it diagnoses errors on labeled development examples, edits only the external skill programs, and retains programs according to held-out validation performance. During \emph{fixed deployment}, the retained programs, label maps, route rules, and selector are frozen and applied to test or real-world inputs without using their labels.

\subsection{External Skill Program}

The editable external state is a skill program
\begin{equation}
    \label{eq:skill_program}
    P=(r,\mathcal{K},\pi).
\end{equation}
The root instruction $r$ specifies the actor's role, the output contract, and the official label ontology. The skill library $\mathcal{K}=\{k_i\}_{i=1}^{K}$ stores localized task knowledge, including fraud cues, audio evidence, fraud-type boundaries, valid-output requirements, and cross-route consistency rules. The route policy $\pi$ determines which skills are active at each decision step. Given route $q$ and the preceding normalized history $h_{q-1}$, it returns an ordered skill list
\begin{equation}
    \label{eq:route_skills}
    K_q
    =
    \pi(q,h_{q-1};\mathcal{K})
    =
    (k_{q,1},\ldots,k_{q,n_q}).
\end{equation}
This representation separates global instructions, local anti-fraud knowledge, and route control. The optimizer may revise any of these external components, while the actor parameters $\theta$ remain unchanged.

\subsection{Offline Skill Optimization}

\paragraph{Trajectory-level error diagnosis.}
Let $\mathcal{D}_{\mathrm{opt}}$ denote the examples used to generate optimization feedback, and let $\mathcal{D}_{\mathrm{prog}}$ be a disjoint held-out split used to select programs. For a candidate program $P$ and an example $d$, FRAUDSkill runs the frozen actor through the ordered routes and records the complete trajectory
\begin{equation}
    \label{eq:trajectory}
    \tau_P(d)
    =
    \operatorname{Run}_{\mathrm{text}}(P,x_d).
\end{equation}
The trajectory contains the raw route responses, provisional parsed labels, route histories, and any skipped decisions. An error analyzer compares the trajectory with the gold labels and returns
\begin{equation}
    \label{eq:error_summary}
    e_P(d)
    =
    E\!\left(\tau_P(d),\mathbf{y}_d\right).
\end{equation}
The error record distinguishes incorrect official labels, unmappable outputs, incorrectly skipped type decisions, cross-route inconsistencies, and systematic errors on minority classes. This trajectory-level diagnosis separates a downstream type error from an upstream fraud decision that prevented the type route from being executed.

\paragraph{External program editing.}
At optimization round $t$, a critic model \cite{zheng2023judging} summarizes the errors observed on a sampled batch $B_t\subset\mathcal{D}_{\mathrm{opt}}$:
\begin{equation}
    \label{eq:textual_feedback}
    g_{t,P}
    =
    G_{\phi}\!\left(
    P,
    \{(d,\tau_P(d),e_P(d)):d\in B_t\}
    \right).
\end{equation}
The feedback $g_{t,P}$ is a natural-language diagnosis rather than a numerical gradient. An editor uses it to produce a bounded neighborhood of revised programs:
\begin{equation}
    \label{eq:neighbor_programs}
    \mathcal{N}(P,g_{t,P})
    =
    \{T_{\psi}(P,g_{t,P},j)\}_{j=1}^{c},
\end{equation}
where $c$ is the branch factor. An edit may revise the root instruction, add or modify a named skill, refine a fraud-type boundary, or update a route-policy rule. The editor never changes the audio-language model or its parameters.

\paragraph{Validation selection and retained program set.}
Starting from a shared initial program $P_0$, FRAUDSkill maintains a beam $\mathcal{B}_t$ of candidate programs. At round $t$, the candidate pool is
\begin{equation}
    \label{eq:candidate_expansion}
    \mathcal{C}_t
    =
    \mathcal{B}_{t-1}
    \cup
    \bigcup_{P\in\mathcal{B}_{t-1}}
    \mathcal{N}(P,g_{t,P}).
\end{equation}
Every candidate is evaluated on $\mathcal{D}_{\mathrm{prog}}$ using the same prespecified route-aware criterion $J$:
\begin{equation}
    \label{eq:frontier_selection}
    \mathcal{B}_t
    =
    \operatorname{Top}_{b}
    \left(
    \mathcal{C}_t;
    J_{\mathcal{D}_{\mathrm{prog}}}
    \right),
\end{equation}
where $b$ is the beam width. Invalid required outputs and incorrectly skipped routes receive zero credit. The rollout, diagnosis, editing, and validation selection steps are repeated for $T$ rounds.The best single program
\begin{equation}
    \label{eq:single_program}
    P^*
    =
    \arg\max_{P\in\cup_{t=0}^{T}\mathcal{B}_t}
    J_{\mathcal{D}_{\mathrm{prog}}}(P),
\end{equation}
defines the text-only variant \emph{FRAUDSkill-Text}. The complete system retains the top $L$ programs according to the same validation criterion:
\begin{equation}
    \label{eq:retained_programs}
    \mathcal{M}
    =
    \operatorname{Top}_{L}
    \left(
    \bigcup_{t=0}^{T}\mathcal{B}_t;
    J_{\mathcal{D}_{\mathrm{prog}}}
    \right)
    =
    \{P^{(1)},\ldots,P^{(L)}\}.
\end{equation}
All retained programs and search settings are frozen before selector fitting and test evaluation. In particular, test performance is not used to choose a program, a seed, or $L$.

\subsection{Fixed Structured Deployment}

\paragraph{Chained execution with canonical history.}
During deployment, every retained program independently guides the same frozen actor. For program $P^{(\ell)}$ and route $q$, FRAUDSkill compiles the root instruction, selected route skills, route query, and preceding normalized history into
\begin{equation}
    \label{eq:compile}
    c_q^{(\ell)}
    =
    C\!\left(
    r^{(\ell)},
    K_q^{(\ell)},
    u_q,
    h_{q-1}^{(\ell)}
    \right).
\end{equation}
The frozen actor then produces an open-ended response
\begin{equation}
    \label{eq:actor_response}
    \widetilde{y}_q^{(\ell)}
    =
    f_{\theta}(x,c_q^{(\ell)}).
\end{equation}
FRAUDSkill applies label projection and route normalization before appending the result to the history. Thus, subsequent routes receive canonical decisions rather than unconstrained natural-language responses.

\paragraph{Label projection.}
For each route, a deterministic projection operator maps the raw response to the official ontology:
\begin{equation}
    \label{eq:canonical_projection}
    y_{q,\mathrm{canon}}^{(\ell)}
    =
    \eta_q\!\left(
    \widetilde{y}_q^{(\ell)};
    \mathcal{Y}_q,\mathcal{A}_q
    \right),
\end{equation}
where $\mathcal{A}_q$ is a fixed map from accepted aliases to official labels. An exact official label is retained, and an unambiguous accepted alias is replaced with its canonical label. If no valid mapping exists, the output is set to $\mathrm{INV}$. The label ontology, alias maps, matching priority, and ambiguity rules are defined and frozen without inspecting test outputs or labels.

\paragraph{Route normalization.}
The route normalizer enforces the conditional protocol:
\begin{equation}
    \label{eq:route_normalization}
    \overline{y}_q^{(\ell)}
    =
    \rho_q\!\left(
    y_{q,\mathrm{canon}}^{(\ell)},
    h_{q-1}^{(\ell)}
    \right).
\end{equation}
The scene and fraud routes are always required. If the normalized fraud decision is \texttt{fraud}, the type route is executed and must return a member of $\mathcal{Y}_{\mathrm{type}}$; an unmappable type response is marked $\mathrm{INV}$. If the normalized fraud decision is \texttt{normal}, the type route is not executed and its value is set to $\mathrm{NA}$. If the fraud decision itself is $\mathrm{INV}$, the downstream type decision is also treated as invalid because its applicability cannot be determined. The normalized decision is appended to the route history:
\begin{equation}
    \label{eq:normalized_history}
    h_q^{(\ell)}
    =
    \operatorname{append}
    \left(
    h_{q-1}^{(\ell)},q,\overline{y}_q^{(\ell)}
    \right).
\end{equation}
Each program therefore produces one normalized trajectory
\begin{equation}
    \label{eq:normalized_trajectory}
    \overline{\mathbf{y}}^{(\ell)}
    =
    \left(
    \overline{y}_{\mathrm{scene}}^{(\ell)},
    \overline{y}_{\mathrm{fraud}}^{(\ell)},
    \overline{y}_{\mathrm{type}}^{(\ell)}
    \right).
\end{equation}

\paragraph{Validation-fitted selection.}
Retained programs may have different route- and class-specific reliability.
FRAUDSkill fits a selector on a calibration split
$\mathcal{D}_{\mathrm{cal}}$ that is disjoint from both test data and the
examples used for textual error feedback. Let
$\overline{\mathbf{Y}}(d)=(\overline{\mathbf{y}}^{(1)}(d),\ldots,
\overline{\mathbf{y}}^{(L)}(d))$ denote the normalized trajectories for example
$d$. The selector family $\Omega_{\mathrm{bal}}$ is a finite, prespecified set
of reliability-weighting and class-balancing configurations. For configuration
$\omega$, route-wise aggregation produces provisional decisions
$S_{\omega,q}$, after which a final projection $\Pi_{\mathcal{C}}$ enforces the
feasible set in Equation~\eqref{eq:feasible_space}:
\begin{equation}
\label{eq:joint_selector}
\begin{aligned}
\widehat{\mathbf{y}}(d;\omega)
&=
\Pi_{\mathcal{C}}
\Bigl(
\bigl\{
S_{\omega,q}
\bigl(
\overline{y}_q^{(1)}(d), \\
&\qquad
\ldots,
\overline{y}_q^{(L)}(d)
\bigr)
\bigr\}_{q\in\mathcal{Q}}
\Bigr).
\end{aligned}
\end{equation}
For compactness, let $\widehat{\mathbf{z}}_{q,\omega}$ and $\mathbf{z}_q$ denote the predicted and gold label vectors for route $q$ on
$\mathcal{D}_{\mathrm{cal}}$. The fitted configuration maximizes
task-averaged Macro-F1 on calibration trajectories:
\begin{equation}
    \label{eq:balanced_selector_fit}
    \omega^*=
    \mathop{\mathrm{arg\,max}}\nolimits_{\omega\in\Omega_{\mathrm{bal}}}
    \frac{1}{|\mathcal{Q}|}\sum_{q\in\mathcal{Q}}
    F_{\mathrm{macro}}^{(q)}
    \left(\widehat{\mathbf{z}}_{q,\omega},\mathbf{z}_q\right).
\end{equation}
At test time, $\omega^*$ is fixed and the final prediction is
\begin{equation}
    \label{eq:full_prediction}
    \widehat{\mathbf{y}}
    =
    \widehat{\mathbf{y}}(x;\omega^*).
\end{equation}
The selector never reads test labels. Class balancing is used only to fit the fixed aggregation rule \cite{dal2015calibrating} and does not change the actor or any retained skill program.

\begin{table*}[!ht]
  \centering
  \scalebox{0.5}{}

  \small
  \setlength{\tabcolsep}{6pt}
  \begin{tabular}{lrrrrrr}
    \toprule
    Method & Macro-F1 & $\Delta$ & W-F1 & Acc. & Joint Acc. & Invalid Rate \\
    \midrule
    Shared baseline  & 41.54 & $\phantom{+}0.00$ & 35.92 & 32.87 &  7.10 & 36.04 \\
    SkillOpt         & 37.67 & $-3.87$            & 36.88 & 33.04 &  5.94 & 29.00 \\
    EvoSkill         & 39.07 & $-2.47$            & 36.79 & 32.67 &  6.16 & 34.19 \\
    FRAUDSkill-Text  & 42.42 & $+0.88$            & 38.87 & 34.88 &  8.18 & 34.48 \\
    \textbf{FRAUDSkill} & \textbf{73.50} & $\mathbf{+31.96}$
                     & \textbf{79.40} & \textbf{78.72} & \textbf{58.87} & \textbf{1.94} \\
    \midrule
    SFT (reference)        & 66.06 & -- & -- & -- & -- & -- \\
    SFT+Memory (reference) & 75.51 & -- & -- & -- & -- & -- \\
    \bottomrule
  \end{tabular}
    \caption{Audio-level results on the complete test set (\%). $\Delta$ denotes
  the absolute Macro-F1 difference from the shared baseline. Methods above the
  divider use the same fixed audio model. SFT and SFT+Memory are contextual
  parameter-training references, and unavailable pipeline metrics are marked
  with ``--''.}
  \label{tab:main_results}
\end{table*}

We use \emph{FRAUDSkill-Text} for the best single external program $P^*$ evaluated without multi-program aggregation. We use \emph{FRAUDSkill} for the complete fixed system consisting of the retained program set $\mathcal{M}$, label projection, route normalization, and validation-fitted selector.

\section{Experiments}
\label{sec:experiments}

We compare external skill adaptation methods under a unified audio model and an
audio-level evaluation protocol. We then isolate the respective roles of
textual program search and structured inference in the complete system.

\subsection{Experimental Setup}
\label{sec:experimental_setup}

\paragraph{Data and protocol.}
After audio-level deduplication, the official SFT split of TeleAntiFraud contains
10,711 training samples and 2,677 test samples, of which 1,453 test samples have
fraud-type annotations. A held-out portion of the training set is used for
program search and selector fitting. All programs, normalization rules, and
selectors are fixed before testing, and no test label is used for their
construction or selection. Evaluation follows the original sequential decision
protocol: the system first recognizes the service scenario, then determines
whether the audio is fraudulent, and predicts a fraud type only if its own
upstream decision is positive. Consequently, an upstream error can prevent a
downstream route from being executed. The evaluation unit is a unique audio
recording. This differs from the 7,021 interaction-level records used in the
original dataset paper, so results under the two protocols are not directly
comparable.

\paragraph{Model.}
All directly compared methods use Qwen2-Audio-7B-Instruct with deterministic
decoding. Its parameters remain fixed throughout adaptation and evaluation.
Differences among these methods therefore arise from how external task knowledge
is represented, optimized, and applied, rather than from changes to the audio
model.

\paragraph{Baselines.}
The shared baseline supplies every method with the same label ontology, output
schema, audio-evidence guidance, and cross-turn consistency rules. SkillOpt
represents this information as a single editable document and revises the skill
through controlled text-space optimization. EvoSkill instead represents it as a
structured skill folder and discovers or modifies local skills in response to
execution failures. These methods instantiate document-level optimization and
modular skill evolution, respectively, but neither explicitly models the
three-stage decision route, projection onto the official label set, or
consistency across routes. FRAUDSkill-Text represents the same task information
as a program comprising a root instruction, named skills, and routing policies;
it isolates the effect of route-aware textual program search. The complete
FRAUDSkill further introduces closed-set projection, route normalization,
complementary trajectories, and class-balanced selection. SFT and SFT+Memory are
included only as parameter-training reference points. Because they update model
parameters, they are not direct baselines and are excluded from the ranking of
external skill methods.

\paragraph{Evaluation metrics.}
Following TeleAntiFraud-Bench, we compute Weighted F1 separately for scenario
recognition, fraud detection, and fraud-type recognition, and report their mean
as W-F1. Because service scenarios and fraud types are imbalanced, the primary
metric is the task-averaged Macro-F1. Accuracy complements the class-balanced
metrics with overall correctness. Joint Accuracy counts a sample as correct only
when every applicable output in the decision chain is correct, while Invalid
Rate measures missing labels, labels outside the official ontology, and outputs
that violate the routing protocol. The original benchmark's LLM-based score for
slow-thinking rationales is not used because FRAUDSkill produces closed-set
decisions rather than free-form reasoning traces for evaluation.

\paragraph{Reporting protocol.}
FRAUDSkill-Text is run with random seeds 42, 43, and 44. The main comparison
reports the arithmetic mean, and variation is measured by the sample standard
deviation. The component analysis starts from the best textual program and adds
structured components cumulatively. Separating these two protocols prevents
search variation from being conflated with the contribution of structured
inference.

\subsection{Main Results}
\label{sec:main_results}

The comparison in Table~\ref{tab:main_results} follows an increasing degree of
task structure. The shared baseline provides fixed task knowledge; SkillOpt and
EvoSkill optimize the external skill artifact; FRAUDSkill-Text adds an explicit
routing structure in the text layer; and the complete FRAUDSkill separates
output constraints and final selection from any single textual program. This
ordering distinguishes generic skill evolution, route-aware program search, and
structured decision control.

Generic skill optimization improves compliance more readily than class-balanced discrimination. The shared baseline obtains a Macro-F1 of 41.54\% with an Invalid Rate of 36.04\%. SkillOpt lowers the Invalid Rate to 29.00\%, yet its Macro-F1 drops to 37.67\%; EvoSkill similarly obtains a 34.19\% Invalid Rate and a Macro-F1 of 39.07\%. Both methods can revise how task information is expressed, but their representations do not explicitly enforce the official label set or dependencies among the three decisions. Better-formed instructions or outputs therefore do not necessarily yield more accurate closed-set classification.

A route-aware program representation partly alleviates this limitation, although textual search alone remains insufficient for final decision making. FRAUDSkill-Text achieves the strongest text-layer result, with an average Macro-F1 of 42.42\%, outperforming the shared baseline by 0.88 percentage points, while improving Joint Accuracy from 7.10\% to 8.18\%. Nevertheless, its Invalid Rate remains high at 34.48\%. These results suggest that organizing root instructions, local skills, and routing policies improves discriminative performance to some extent, but a single textual program must still perform both open-ended generation and structured decision control.

The complete FRAUDSkill separates these responsibilities across different components. Closed-set projection and route normalization convert open-ended generations into protocol-valid candidates, complementary program trajectories provide alternative decisions, and a validation-fitted selector produces the final output. This design raises Macro-F1 to 73.50\%, an absolute improvement of 31.96 percentage points over the shared baseline. W-F1, Accuracy, and Joint Accuracy reach 79.40\%, 78.72\%, and 58.87\%, respectively, while the Invalid Rate falls to 1.94\%. The substantial gain observed after introducing structured inference suggests that closed-set control, route consistency, and class-aware selection contribute more directly to the final detection outcome than further skill rewriting.
The SFT and SFT+Memory rows are parameter-updating references, not frozen-skill baselines.

\subsection{Ablation Studies}
\label{sec:ablation_studies}

\begin{figure}[t]
  \centering
  \IfFileExists{figures/figure_experiments.png}{%
    \includegraphics[width=\columnwidth]{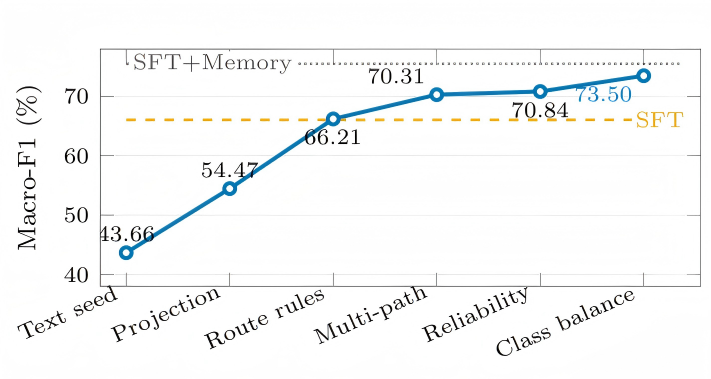}%
  }{%
    \fbox{\parbox[c][4.2cm][c]{0.94\columnwidth}{\centering Placeholder for Figure 2: text-layer comparison and cumulative structured inference analysis.}}%
  }
  \caption{Compact summary of main results. The curve shows cumulative gains from stacking structured components on the best text seed; SFT and SFT+Memory are external reference lines.}
  \label{fig:structured_results}
  \vspace{-2mm}
\end{figure}

The preceding analyses identify two distinct limitations: variation across
single-program searches and a gap between output validity and class
discrimination. Starting from the best text-layer program, Figure~\ref{fig:structured_results} shows how the complete system addresses them in sequence. Closed-set projection
first raises Macro-F1 from 43.66\% to 54.47\%, and route normalization further raises
it to 66.21\%. Together, label projection and decision-chain consistency account
for 22.55 percentage points, suggesting that many text-layer outputs contain relevant
semantics but either fail to map to an official label or conflict with an
upstream decision.

Once candidate trajectories share a valid representation, complementary multi-path inference raises Macro-F1 from 66.21\% to 70.31\%. Different programs make complementary errors across routes and classes, making the retained trajectory set more robust than relying on a single program. Reliability weighting further raises Macro-F1 to 70.84\%, outperforming majority voting by 0.53 percentage points. Adding class-balanced selection produces the final Macro-F1 of 73.50\%, yielding an additional gain of 2.66 percentage points. These results suggest that the main benefit of the selection stage lies in protecting minority classes \cite{han2025mitigating}, rather than simply reinforcing the majority decision.

The structured components contribute 29.84\% Macro-F1 points over the best textual program. Closed-set projection and route normalization construct valid, consistent candidates; multi-path inference exploits complementarity among programs; and class-balanced selection performs the final discrimination.
This division of labor turns text-layer skill search from single-program optimization into a structured decision process for closed-set audio anti-fraud detection.

\begin{figure*}[t]
  \centering
  \IfFileExists{figures/figue_case_study.png}{%
    \includegraphics[width=\textwidth]{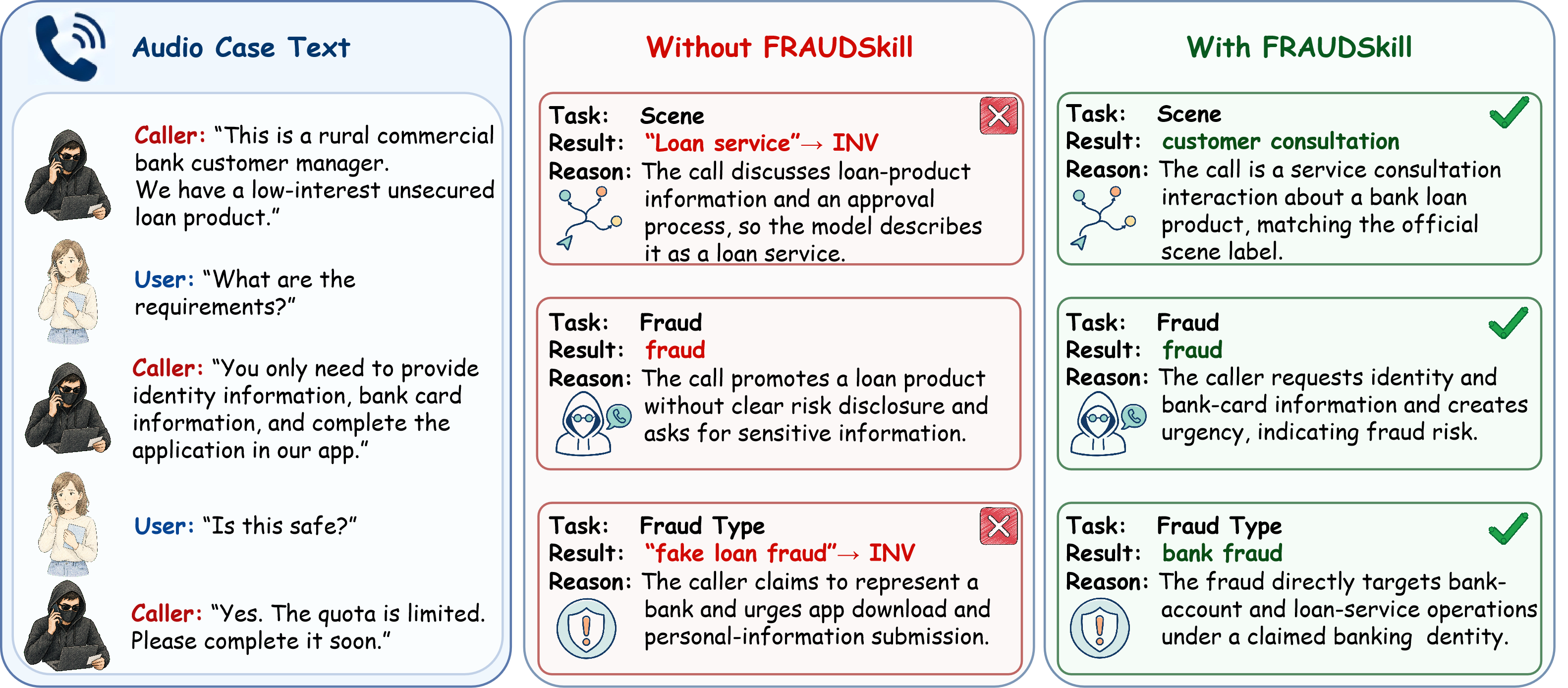}%
  }{%
    \fbox{\parbox[c][5.2cm][c]{0.96\textwidth}{\centering Placeholder for the FRAUDSkill case study figure.}}%
  }
  \caption{Case study of structured skill adaptation on a loan-related fraud call. The left column shows the audio case text. Without FRAUDSkill, the frozen audio actor provides plausible reasons but produces non-ontology labels such as "loan service" and "fake loan fraud", which are mapped to \textsc{INV}. With FRAUDSkill, external skill artifacts guide the same frozen actor toward protocol-compatible closed-set outputs with route-level reasons, yielding the official labels customer consultation, fraud, and bank fraud.}
  \label{fig:case-study}
  \vspace{-2mm}
\end{figure*}

\begin{table}[h]
\centering
\small
\setlength{\tabcolsep}{4pt}

\begin{tabular}{llr}
\toprule
Route & Diagnostic & Count \\
\midrule
Scene & evaluated examples & 2,677 \\
Scene & all errors & 1,985 \\
Scene & missing/off-ontology & 1,808 \\
\midrule
Fraud & evaluated examples & 2,677 \\
Fraud & all errors & 881 \\
Fraud & false normal & 529 \\
Fraud & false fraud & 340 \\
\midrule
Type & annotated examples & 1,453 \\
Type & wrong annotated cases & 1,208 \\
\bottomrule
\end{tabular}
\caption{Test-set error counts of FRAUDSkill-Text.}
\label{tab:error_counts}
\vspace{-2mm}
\end{table}

\section{Error Analysis}
As shown in Table~\ref{tab:error_counts}, FRAUDSkill-Text exhibits errors across all decision routes. The scene-level error rate reaches 74.2\%, with missing or off-ontology outputs accounting for 91.1\% of these errors, indicating difficulty in converting open-ended generations into valid closed-set labels. Although the fraud-route error rate is lower at 32.9\%, false-normal decisions account for 60.0\% of its errors and can block downstream fraud-type classification; the type route remains the most difficult, with an 83.1\% error rate on annotated examples. These findings motivate the use of label projection, route normalization, and complementary multi-path inference to improve label validity and cross-route consistency.

\section{Case Study}

Figure 4 compares FRAUDSkill on a typical "bank low-interest loan impersonation" scam call. Without  FRAUDSkill, the frozen model tends to generate open-ended labels. It describes the scene as "Loan service" and the fraud type as "fake loan fraud." These terms are semantically related to the call but are not included in the predefined label set, so they are mapped to invalid outputs. As a result, only the fraud judgment is valid, giving one valid prediction out of three. With FRAUDSkill, label projection and route normalization constrain the outputs to the official label space. The same call is judged as customer consultation, fraud, and bank fraud, and all three predictions are valid. This sample-level change is consistent with the global decrease in invalid outputs from 36.04\% to 1.94\%. The rationales also become more specific and better aligned with the label definitions. Instead of saying only that the caller promotes a loan without disclosing risks, the model points to cues such as requesting identity and bank-card information and creating urgency through a limited quota. This shows that FRAUDSkill does not simply shorten free text. It keeps the original meaning while guiding it into a controlled label space. Since the Scene, Fraud, and Fraud Type decisions must be checked step by step, FRAUDSkill makes each decision and its reason inspectable, so errors can be located and traced. These improvements are obtained without updating model weights or task-specific tuning, offering a training-free and plug-and-play way to deploy frozen audio-language models for trustworthy anti-fraud decisions.

\section{Conclusion}

We present FRAUDSkill, a structured frozen-weight ASO method for audio anti-fraud detection. It addresses the mismatch between open-ended audio-language generation and closed-set fraud deployment by adapting external skill artifacts rather than audio-model parameters. The text layer extends APO-style search from one prompt to a routed skill program. The inference layer then normalizes complementary frozen-actor trajectories and applies class-balanced reliability selection.

The results support two scoped conclusions. First, text-layer skill search alone is useful but unstable in this setting: FRAUDSkill-Text marginally improved mean Macro-F1 over the shared baseline, with variation larger than the mean gain. Second, the complete structured system is much stronger. Under the studied protocol, it achieved competitive performance without updating the audio actor, clearly above the plain SFT reference and slightly below SFT+Memory. These comparisons are protocol-bound and do not imply a task independent ranking. They show that, for closed-set audio anti-fraud deployment, validity control, complementary trajectories, and class-aware selection are central to making frozen-weight adaptation competitive with weight-updating references.
\section{Ethical Statement}
FRAUDSkill is intended only for defensive research on
telecom-fraud detection. Its synthetic, non-private artifacts
must support only defensive analysis and never deception,
impersonation, or scam training.

\begingroup
\footnotesize
\bibliography{references}
\endgroup
\clearpage
\appendix
\section*{Supplementary Material}
\setcounter{table}{0}
\renewcommand{\thetable}{S\arabic{table}}
\section{Experimental Protocol Details}

This supplement records protocol and implementation details for reproducibility. It does not repeat the main-paper method description, framework figure, case study, or headline result discussion.

\paragraph{Task protocol.}
Each audio example is evaluated through an ordered route sequence: scene, fraud, and type. The scene route predicts one service scenario from the official scene ontology. The fraud route predicts either \texttt{normal} or \texttt{fraud}. The type route is evaluated only when fraud is predicted or annotated as \texttt{fraud}; otherwise its value is \texttt{NA}. A required output that is missing, malformed, or not recoverable as an official label is mapped to \texttt{INV}. This distinguishes an unusable output from a valid negative decision.

\paragraph{Official label spaces.}
The scene labels used by the released evaluation configuration are \texttt{account support}, \texttt{customer consultation}, \texttt{e-commerce promotion}, \texttt{public service}, \texttt{delivery service}, and \texttt{other}. The fraud labels are \texttt{normal} and \texttt{fraud}. The fraud-type labels are \texttt{bank fraud}, \texttt{phishing}, \texttt{identity theft}, \texttt{investment fraud}, \texttt{romance fraud}, \texttt{fake customer service}, \texttt{lottery fraud}, and \texttt{other fraud}.

\paragraph{Metric computation.}
The primary metric is task-averaged Macro-F1 over the three routes. The paper also reports weighted F1, accuracy, joint accuracy, and invalid rate. Joint accuracy requires every applicable route in the chain to be correct. Invalid rate measures route outputs that are malformed, missing, or outside the official label space after parsing and normalization.

\paragraph{Split isolation.}
Textual skill search uses training examples for candidate generation and held-out validation examples for program selection. The validation-fitted selector is calibrated on a validation split disjoint from the test set and from the examples used to generate textual error feedback. Test labels are not used by the skill editor, label projection, route normalization, or selector fitting.

\section{External Skill Search Settings}

\paragraph{Frozen actor.}
All frozen-weight methods use the same fixed audio-language actor. The actor listens to the audio input and generates route-level text. FRAUDSkill changes only external artifacts: root instructions, named skills, route policies, label maps, deterministic route normalizers, and selector configurations.

\paragraph{Search configuration.}
The released configuration uses seed 42 for the default search trace, beam width 3, branch factor 3, five search rounds, and optimization batches of 64 examples. Candidate programs are scored on held-out validation data using a route-aware score during search. The reported paper metric remains task-averaged Macro-F1.

\paragraph{Critic and editor interface.}
For each candidate program, the frozen actor is rolled out on a batch of examples. The trajectory records raw route text, parsed route decisions, normalized decisions, route histories, and route-level errors. A textual critic summarizes recurring failure modes, such as invalid labels, off-route predictions, skipped required routes, and cross-route inconsistencies. A typed editor then proposes revised external programs. The editor may revise the root instruction, named skills, local skill content, or route policy, but it does not update the audio-model parameters.

\paragraph{Final artifact selection.}
FRAUDSkill-Text denotes the best single selected external program. The full FRAUDSkill system retains multiple complementary programs and fixes the closed-set projection, route normalizer, and class-balanced reliability selector before test evaluation.

\section{Text-Search Stability and Repair Results}

\paragraph{Text-search stability.}
Table~\ref{tab:supp_text_seed_stability} reports the test-set results of FRAUDSkill-Text across random seeds. This table characterizes stochastic variation in the text-level skill search. The mean Macro-F1 is $42.42\%$, with a sample standard deviation of $1.39$ points across the three runs.

\begin{table}[H]
\centering
\small
\begin{tabular}{lccccc}
\toprule
Run & Macro & W-F1 & Acc. & Joint & Invalid \\
\midrule
Seed 42 & 43.66 & 40.66 & 36.60 & 8.55 & 30.77 \\
Seed 43 & 42.69 & 41.43 & 36.83 & 9.08 & 33.14 \\
Seed 44 & 40.92 & 34.52 & 31.22 & 6.91 & 39.53 \\
\midrule
Mean & 42.42 & 38.87 & 34.88 & 8.18 & 34.48 \\
Sample std. & 1.39 & 3.79 & 3.17 & 1.13 & 4.53 \\
\bottomrule
\end{tabular}
\caption{Test-set results of FRAUDSkill-Text across random seeds (\%). The standard deviation is the sample standard deviation.}
\label{tab:supp_text_seed_stability}
\end{table}

\paragraph{Text-level repair results.}
Table~\ref{tab:supp_text_repairs} reports validation-set results for text-level output repairs. The repairs improve validity-oriented metrics, especially invalid rate and joint accuracy, but they do not by themselves produce the final structured-inference gains reported in the main paper.

\begin{table}[H]
\centering
\small
\begin{tabular}{lccccc}
\toprule
Repair & Macro & W-F1 & Acc. & Joint & Invalid \\
\midrule
Original & 41.67 & 39.93 & 35.82 & 8.08 & 30.75 \\
Strict scene & 40.14 & 38.55 & 34.01 & 7.24 & 30.65 \\
Type boundary & 40.26 & 41.33 & 36.50 & 8.96 & 28.07 \\
Balanced & 41.18 & 42.35 & 37.86 & 9.62 & 25.28 \\
\bottomrule
\end{tabular}
\caption{Validation-set results for text-level output repairs (\%). Textual constraints improve validity and joint accuracy, but they do not by themselves produce the final structured-inference gains.}
\label{tab:supp_text_repairs}
\end{table}

\section{Reproducibility Details}

\paragraph{Code and data artifact.}
The code and data supplement contains the evaluation scripts, configuration files, official label ontology, result tables, prediction trace schemas, selected skill-program artifacts, selector configuration, and local replay scripts. The smallest verification path regenerates the paper-table CSV files from released JSON result files and checks that prediction traces conform to the released schema.

\paragraph{Local replay.}
The artifact provides a local replay mode for checking the search and evaluation pipeline without requiring private API credentials or the full audio benchmark. This mode verifies the route sequence, trace schema, program-selection bookkeeping, deterministic normalization logic, and output file structure. A full rerun of the original experiment requires access to the TeleAntiFraud benchmark audio split, the frozen audio-language model, and the critic/editor model interfaces used for external skill search.

\paragraph{Unavailable components.}
Raw audio files and frozen model weights are not duplicated in the supplement. The raw audio benchmark is governed by the source dataset release and its access conditions, while model weights follow the upstream model provider's license and distribution channel. API keys and service credentials are intentionally excluded. The research artifact instead records model-interface variables, route prompts, result schemas, selected external skills, and replayable post-processing logic.

\paragraph{Randomness and compute.}
The text-level search is stochastic because candidate generation depends on sampled batches and textual critic/editor outputs. The paper reports FRAUDSkill-Text variation over seeds 42, 43, and 44. The complete system uses retained programs and selector configurations fixed before test evaluation. Compute disclosure is separated into frozen actor inference, external critic/editor calls during skill search, and deterministic post-processing during structured deployment.

\end{document}